\documentclass[a4paper, 10pt, twocolumn]{article}
\usepackage[T1]{fontenc}

\usepackage{amsfonts,amssymb,amsmath,amsthm}
\usepackage{subfigure}
\usepackage{graphicx}
\usepackage{color}
\usepackage{textcomp}
\usepackage{soul}
\usepackage[footnotesize]{caption}
\usepackage[ruled,vlined]{algorithm2e}

\usepackage{titlesec}
\titlespacing{\paragraph}{%
  0pt}{
  0.4\baselineskip}{
  1em}
\titlespacing{\section}{%
  0pt}{
  0.4\baselineskip}{
  0.2\baselineskip}

\usepackage[top=1.5cm, left=1.5cm, right=1.5cm, bottom=1.5cm]{geometry}

\newcommand{\cel}{{\sf c}}

\newcommand{\bs}{\boldsymbol}
\newcommand{\bb}{\mathbb}
\newcommand{\cl}{\mathcal}

\newcommand{\ts}{\textstyle}
\newcommand{\ie}{\emph{i.e.}, }

\newcommand{\tiid}{%
    \ifmmode
        \mathrm{iid}%
    \else%
        iid\xspace%
    \fi%
}

\newcommand{\bcl}[1]{\boldsymbol{\mathcal{#1}}}
\newcommand{\tm}{\times}
\newcommand{\otm}{\otimes}

\newcommand{\jj}{\mathrm{j}}

\newcommand{\ong}{\emph{on-the-grid} }
\newcommand{\ofg}{\emph{off-the-grid} }

\newcommand{\amin}[1]{\underset{\hspace{1mm}#1\hspace{1mm}}{\text{arg\,min}\hspace{1mm}}}

\newcommand{\bpsi}{\bs \psi}

\newcommand{\OP}{\Omega_{\cl P}}

\newcommand{\oml}[1]{\omega^{#1}_{n_{#1}}}
\newcommand{\OPl}[1]{\Omega_{\cl P_{#1}}}

\definecolor{darkgreen}{rgb}{0.1,0.6,0.2}

\title{\Large{\textbf{Factorization over interpolation: A fast continuous orthogonal matching pursuit}}}

\author{Gilles Monnoyer de Galland$^{1,2}$, Luc Vandendorpe$^1$ and Laurent
Jacques$^2$.\\
\footnotesize $^1$CoSy.\ $^2$ISPGroup.\
ICTEAM/ELEN, UCLouvain, Louvain-la-Neuve, Belgium\thanks{GM and LJ are funded by the Belgian FNRS.}. \vspace{-3mm}
}
\date{\empty} 

\renewenvironment{abstract}{\bf\small {\em\ Abstract---}}{}
\titleformat*{\section}{\large\bfseries}

\begin{document}

\maketitle

\begin{abstract} 
We propose a fast greedy algorithm to compute sparse representations of signals from \emph{continuous} dictionaries 
that are \emph{factorizable}, \ie with atoms that can be separated as a product of sub-atoms.
Existing algorithms strongly reduce the computational complexity of the sparse decomposition of signals in \emph{discrete} factorizable dictionaries. 
On another flavour, existing greedy algorithms use interpolation strategies from a discretization of continuous dictionaries to perform \ofg decomposition.
Our algorithm aims to combine the factorization and the interpolation concepts to enable low complexity computation of continuous sparse representation of signals. 
The efficiency of our algorithm is highlighted by simulations of its application to a radar system.
\end{abstract}

\section{Introduction}
\label{sec_intro}
Computation of sparse representations is beneficial in many applicative fields such as radar signal processing, communication or remote sensing~\cite{baraniuk2007,zheng2017,YD_zhang2013}. 
This computation is based on the assumption that a signal $\bs y$ decomposes as a linear combination of a few atoms taken from a \emph{dictionary} $\cl D$.
In this paper, we focus on continuous parametric dictionaries~\cite{gribonval2001, figueras2006, jacques2008} which associate each parameter $\bs p$ from the continuous parameter set $\cl P$ to an atom $\bs a(\bs p)\in\bb C^M$. 
Thereby, the decomposition of $\bs y\in\bb C^M$ reads
\begin{equation}
    \ts \bs y = \sum_{k=1}^K\alpha_k \bs a(\bs p_k),
    \label{eq_y}
\end{equation}
where for all $k\in[K]$, $\alpha_k\in\bb C$ and $\bs p_k\in\cl P$.

In some applications~\cite{winkler2007, lutz2014, feuillen2016}, the atoms $\bs a(\bs p_k)$ factorize as a product of sub-atoms, each depending on a distinct set of components of $\bs p_k$.
In that case, greedy algorithms such as presented in \cite{zubair2013,fang2015} can leverage this property to strongly reduce the computational complexity of the decomposition.

These approaches, however, capitalize a discretization of $\cl P$ and assume that the parameters $\{\bs p_k\}_{k=1}^K$ match the resulting grid~\cite{candes2008}.
Yet, estimations of parameters from such discretized models are affected by grid errors~\cite{azodi2016}. 
Although a denser grid reduces this effect, it tremendously increases the dimensionality of the problem to solve. 
Continuous reconstruction algorithms do not require such dense grids as they perform \ofg estimations of parameters~\cite{tang2013, mishra2014, traonmilin2018}.
In~\cite{simoncelli2011}, a continuous version of the Basis Pursuit is derived from the construction of an interpolated model that approximates the atoms. 
In~\cite{knudson2014}, the authors similarly designed the Continuous OMP (COMP) from the same interpolation concept. 

We propose a Factorized COMP (F-COMP) that combines the concepts of interpolation and factorization to enable a fast and accurate reconstruction of sparse signals.
We applied our algorithm to the estimation of the ranges and speeds of targets using a Frequency Modulated Continuous Wave (FMCW) radar.
Simulations validated the superiority of using low-density grids with \ofg algorithms instead of denser grids with \ong algorithms. 

\paragraph*{Notations:} Matrices and vectors are denoted by bold uppercase and lowercase symbols, respectively. The tensors are denoted with bold calligraphic uppercase letters. 
The outer product is $\otm$, 
$\small\|\cdot\small\|_F$ is the Frobenus norm, $[N] := \{1,\cdots,N\}$, $\jj=\sqrt{-1}$, and $\cel$ is the speed of light.


\section{Problem Statement}
\label{sec_prob}
We consider the problem of estimating the values of $K$ parameters $\{\bs p_k \}_{k=1}^K \subset \cl P$ from a measurement vector $\bs y \in \bb C^M$.
This measurement is assumed to decomposes as \eqref{eq_y}, with $K\ll M$.
The parameters are known to lie in a separable parameter domain $\cl P \subset \bb R^L$ such that $\cl P := \cl P_1 \tm \cdots \tm \cl P_L$ with $\cl P_\ell \subset \bb R$ for each $\ell\in[L]$.
For all $k\in[K]$, $\bs p_k$ decomposes in
\begin{equation}
    \bs p_k := (p_{k,1}, \cdots, p_{k,L})^\top,
\end{equation}
with $p_{k,\ell} \in \cl P_\ell$ for all $\ell\in[L]$. In \eqref{eq_y} the atoms $\bs a(\bs p_k)$ are taken from a continuous dictionary defined by $\cl D := \{\bs a(\bs p) : \bs p\in \cl P\}$. 

In this paper, we consider the particular case of dictionaries of atoms that factorize in $L$ sub-atoms.
More precisely, introducing the tensor $\bcl A(\bs p)\in\bb C^{M_1\tm\cdots\tm M_L}$ reshaping $\bs a(\bs p)\in\bb C^M$
\begin{equation}
    \cl A_{m_1,m_2,\cdots m_L}(\bs p) := a_{\Bar{m}}(\bs p),
    \label{eq_A2a}
\end{equation}
with $\Bar{m} := m_L+\sum_{\ell=1}^{L-1}(m_\ell -1) \prod_{i=\ell+1}^LM_i \in [M]$, $m_\ell\in[M_\ell]$ for all $\ell\in[L]$ and $M = M_1M_2\cdots M_L$, we assume that the atom $\bcl A(\bs p_k)$ decomposes in
\begin{equation}
    \bcl A(\bs p_k) := \bpsi_1(p_{k,1})\otm \cdots \otm \bpsi_L(p_{k,L}).
    \label{eq_A_fact}
\end{equation}
In \eqref{eq_A_fact}, each $\bpsi_\ell(p_{k,\ell}) \in \bb C^{M_\ell}$ is a sub-atom taken from the continuous dictionary $\cl D_\ell := \{\bpsi_\ell(p) : p\in\cl P_\ell\}$. 
In the tensor reshaped domain, the decomposition \eqref{eq_y} becomes
\begin{equation}
    \ts \bcl Y = \sum_{k=1}^K \alpha_k \bcl A(\bs p_k),
    \label{eq_Y_fact}
\end{equation}
where $\bcl Y\in\bb C^{M_1\tm\cdots\tm M_L}$ is the tensor-shaped measurement, \ie $\cl Y_{m_1,m_2,\cdots, m_l} := y_{\Bar{m}}$.

Recovering the parameters $\{\bs p_k\}_{k=1}^K$ from the factorized model \eqref{eq_Y_fact} can be made fast. 
For instance, in work \cite{zubair2013,fang2015}, the authors consider an adaptation of OMP, that we coin Factorized OMP (F-OMP), which leverages the decomposition \eqref{eq_A_fact} to reduce the dimensionality of the recovery problem.
Yet, F-OMP only enables the estimation of \ong parameters taken from a finite discrete set of parameters.
In the next section, we build a model based on the same grid which enables the greedy estimation of \ofg parameters  while similarly leveraging the factorization.

\section{Factorization over Interpolation}
\label{sec_fact}
From the general non-factorized model \eqref{eq_y}, the algorithm Continuous OMP (COMP) \cite{knudson2014} extends OMP and succeeds to greedily estimate \ofg parameters. 
COMP operates with a parameter grid which results from the sampling of $\cl P$. 
The atoms of the continuous dictionary $\cl D$ are approximated by a linear combination a multiple atoms which are defined from the grid. 
This combination enables to interpolate (from the grid) the atoms of $\cl D$ that are parameterized from \ofg parameters.
Our algorithm F-COMP applies the same interpolation concept to the atoms $\bcl A(\bs p)$, which are factorized by \eqref{eq_A_fact}.

Let us define the separable grid $\OP \subset \cl P$ such that $\OP = \OPl{1} \tm \cdots \tm \OPl{L}$, with $\OPl{\ell} := \{\oml{\ell}\}_{n_\ell=1}^{N_\ell}\subset \cl P_\ell$ for all $\ell\in[L]$.
We propose a ``factorization over interpolation" strategy where each \ofg atom $\bcl A(\bs p_k)$ is interpolated by $I$ \ong atoms 
\begin{equation}
    \ts \bcl A(\bs p_k) \simeq \sum_{i=1}^I c_k^{(i)}\bcl A^{(i)}[\bs n(k)].
    \label{eq_interp_fact}
\end{equation}
In \eqref{eq_interp_fact}, the indices $\bs n(k) := (n_1(k), \cdots, n_L(k))$ depend on the interpolation scheme and on $\bs p_k$, and each $\bcl A^{(i)}[\bs n(k)]$ is the $i$-th interpolation atom associated to the $\bs n(k)$-th element of the grid $\OP$.
The coefficients $c_k^{(i)}$ are obtained from 
\begin{equation}
    (c_k^{(1)}, \cdots c_k^{(I)}) = \cl C_{\bs n(k)}(\bs p_k),
\end{equation}
where $\cl C_{\bs n(k)}(\bs p)$ is a function defined from the choice of interpolation pattern \cite{simoncelli2011,knudson2014}.
In this scheme, for all $i\in[I]$, we decompose the global interpolation atoms $\bcl A^{(i)}[\bs n(k)]$ using interpolation sub-atoms denoted by $\bpsi_\ell^{(i)}[n_\ell(k)]$, \ie
\begin{equation}
    \bcl A^{(i)}[\bs n(k)] = \bpsi_1^{(i)}[n_1(k)] \otm \cdots \otm \bpsi_L^{(i)}[n_L(k)].
    \label{eq_interp_fact2}
\end{equation}

The factorization \eqref{eq_interp_fact2} is enabled by the properties of the interpolant dictionaries.
It is for instance the case for the dictionaries describing FMCW chirp-modulated radar signals we detail in Sec.~\ref{sec_rad}. 
From such signals, we can efficiently estimate \ofg values of $\{\bs p_k\}_{k=1}^K$ using the Factorized Continuous OMP that we explain in the next section.

\section{Factorized Continuous OMP}
\begin{algorithm}[b!]
\SetKwInOut{Input}{Input}\SetKwInOut{Output}{Output}
\SetKwFor{While}{While}{:}{end}

\Input{$K$, $\bcl Y$, $\big\{\bcl A^{(i)}[\bs n]\big\}_{(i,\bs n) \in [I]\tm \cl N}$, $\OP$.}

\Output{$\{\hat{\alpha}_k\}_{k=1}^K, \{\bs p_k)\}_{k=1}^K$} 

\Begin{

 Initialization: $\bcl{R}^{(1)} = \bcl{Y}$, $\Omega = \emptyset$;

\While{$k \leq K$}{
    \vspace{-5mm}
    
    \begin{equation}
    \!\!\!\!\!\!\!\hat{\bs n}(k) = \amin{\bs n\in\cl N}\big(\underset{\bs \beta\in\bb C^I}{\text{min}} \big\|\sum_{i=1}^I\beta_i \bcl A^{(i)}[\bs n]-\bcl R^{(k)}\big\|_F^2 \big) \hspace{1.5mm} \text{(9)} \nonumber \setcounter{equation}{9}
    \label{eq_indsel_alg}
    \end{equation}
    
    \vspace{-4mm}
    
    $$\!\!\!\!\!\!\!\!\!\!\!\!\!\!\!\!\!\!\!\!\!\!\!\!\!\!\!\!\!\!\!\!\!\!\!\!\!\!\!\!\!\!\!\!\!\!\!\!\!\!\!\!\!\!\!\!\!\!\!\!\!\!\!\!\!\!\!\!\!\!\!\!\!\!\!\!\!\!\!\!\!\!\!\!\!\Omega \leftarrow \Omega \cup \{\hat{\bs n}(k)\}$$
    
    \vspace{-9mm}
    $$\!\!\!\!\!\!\!\!\! \big\{\hat{\bs \beta}_{k'}\big\}_{k'=1}^k \hspace{-1mm} = \hspace{-3mm}\amin{\{\hat{\bs \beta}_{k'}\in \bb C^I\}_{k'=1}^k}\hspace{-2.5mm} \small{\Big\|\hspace{-0.5mm} \sum_{k'=1}^k \sum_{i=1}^I \beta_{k'}^{(i)} \bcl A^{(i)}\big[\hat{\bs n}(k')\big] - \bcl Y \Big\|_F^2} $$
    
    \vspace{-2mm}
    
     $\!\!\!\!\bs r^{(k+1)} = \bs y - \sum_{k'=1}^k \sum_{i=1}^I \hat \beta_{k'}^{(i)} \bcl A^{(i)}\big[ \hat {\bs n}(k')\big]$

     $k \leftarrow k+1$
}
\vspace{2mm}
    \hspace{-2mm}for all $k\in[K]$,\\
    \vspace{-5mm}
    
    \begin{equation}
        \!\!\!(\hat{\alpha}_k, \hat{\bs p}_k) = \hspace{-1mm}\amin{\alpha\in\bb C, \bs p\in\cl P}\hspace{-2mm} \big\| \alpha \cl C_{\hat {\bs n}(k)}(\bs p) - \hat{\bs \beta}_k \big\|_2^2. 
        \label{eq_corr_step}
    \end{equation}
    \vspace{-3mm}
}
  \caption{Factorized Continuous OMP (F-COMP)}
  \label{alg_COMP}
\end{algorithm}
Alg.~\ref{alg_COMP} formulates F-COMP for a generic interpolation scheme.
F-COMP leverages the factorized interpolated model \eqref{eq_interp_fact} to estimate \ofg parameters with a reduced complexity with respect to COMP \cite{knudson2014}. 
It follows the same steps as COMP and greedily minimizes $\big\|\bcl Y - \sum_{k=1}^K\alpha_k\sum_{i=1}^I c^{(i)}_k \bcl A^{(i)}[n(k)]\big\|_F^2$.
In Alg.~\ref{alg_COMP}, we use $\cl N$ to denote $[N_1] \tm \cdots \tm [N_L]$ and $\hat{\bs \beta}_k := (\hat \beta^{(1)}_1, \cdots \hat \beta^{(I)}_k)$, where $\hat \beta_k^{(i)}$ estimates $\alpha_k c^{(i)}_k$.

The decomposition of the atoms $\bcl A(\bs p_k)$ enables to compute the step in \eqref{eq_indsel_alg} with a complexity $O(I N_1\cdots N_L\min_{\ell\in[L]}(M_\ell))$ instead of $O(I N_1\cdots N_L M_1\cdots M_L)$ in COMP.
This is achieved by extending the methodology of F-OMP \cite{fang2015} to our interpolation-based model.

\section{Application to Radars}
\label{sec_rad}
We applied F-COMP for the estimation of ranges $\{r_k\}_{k=1}^K$ and radial speeds $\{v_k\}_{k=1}^K$ of $K$ point targets using a Frequency Modulated Continuous Wave (FMCW) radar that emits chirp-modulated waveforms.
The model is similar to the one provided in \cite{feuillen2016}.
The received radar signal is coherently demodulated and sampled with a regular sampling rate $1/T_s$, $M_s$ samples acquired per chirp, and $M_c$ chirps are acquired. 
With a few assumptions on the radar system \cite{feuillen2016, liu2010, bao2014}, the resulting sampled measurement vector $\bs y \in \bb C^{M_cM_s}$ is approximated for $m_s\in[M_s]$, and $m_c\in[M_c]$ by  
\begin{equation}
    y_{m_cM_s+m_s} \simeq \sum_{k=1}^K\alpha_k e^{-\jj2\pi\frac{B}{M_s}\frac{2r_k'}{\cel} m_s} e^{-\jj2\pi f_0M_sT_s \frac{2 v_k}{\cel} m_c},
    \label{radar_eq}
\end{equation}
where $r_k' = r_k+\frac{f_0M_sT_s}{B}v_k$. In \eqref{radar_eq}, $B$ and $f_0$ respectively are the bandwidth and the carrier frequency of the transmitted waveform.
Given \eqref{radar_eq}, the measurement $\bs y$ is reshaped as explained in Sec.~\ref{sec_prob} and expressed by \eqref{eq_Y_fact} and \eqref{eq_A_fact} where $L=2$, 
\begin{align}
    (\bpsi_1(r_k'))_{m_s} &:= e^{-\jj2\pi\frac{B}{M_s}\frac{2r_k'}{\cel}m_s},\\
    (\bpsi_2(v_k))_{m_c} &:= e^{-\jj2\pi f_0M_sT_s \frac{2 v_k}{\cel} m_c}.
\end{align}
In our application, we used an order-1 Taylor interpolation such as explained in~\cite{simoncelli2011, knudson2014} to implement \eqref{eq_interp_fact} and \eqref{eq_interp_fact2}.

\begin{figure}[t!]
    \centering
    \includegraphics[width=\linewidth]{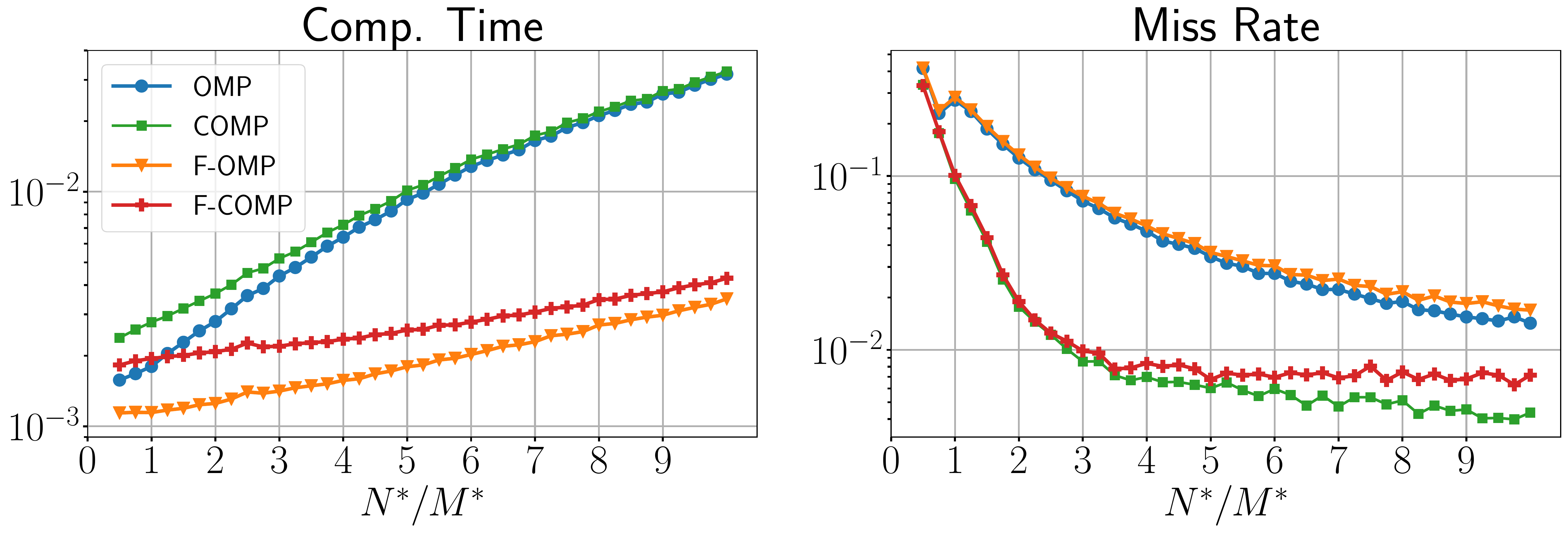}
    \caption{Comparison of (a) Computation Time and (b) Miss Rate of (F)(C)OMP in function of the number of bins the location and in the velocity grids ($N^* = N_1 = N_2$). The simulated system has $M^*=M_1=M_2=16$. $B = 200$MHz, $f_0=24$GHz, $T_s=5\mu$s and $T_c = M_sT_s$. Each dot is obtained by averaging the values resulting from 10,000 realisations of random sets of $K=5$ independent targets.}
    \label{fig}
\end{figure}

The radar model \eqref{radar_eq} is an approximation of the exact radar signal that enables the use of F-COMP. 
With the exact model, we can use COMP and expect more accurate estimation of ranges and speeds with a higher computation time than F-COMP.
This is observed in Fig.~\ref{fig}, where an estimation is a \emph{miss} when $\sqrt{\big(\frac{\hat r_k - r_k}{\cel/2B}\big)^2 + \big(\frac{\hat v_k - v_k}{\cel/(4f_0M_cT_c)}\big)^2} < 1$.
The factorized algorithms (F-OMP and F-COMP) are faster but miss more often the estimation than their non-factorized counterparts (OMP and COMP). 
The continuous algorithms have a lower miss rate because they are not affected by the grid errors.
F-COMP appears as the best trade-off between performance and computation time for most values of MR it can reach.

\section{Conclusion}
In this work, we designed the Factorized Continuous OMP which leverages the factorized structure of dictionaries to efficiently compute continuous sparse representations. 
We proposed an implementation of the algorithm for a practical radar application.
Although this implementation remains simple, our simulations showed F-COMP as the best trade-off between performance and compuation time.
In future work, we may investigate the extension of more sophisticated, and of higher order, interpolation schemes to factorizable dictionaries.

\appendix
\bibliographystyle{unsrt}
\begin{small}
\bibliography{main.bib}

\begin{thebibliography}{10}

\bibitem{baraniuk2007}
R.~{Baraniuk} and P.~{Steeghs}.
\newblock Compressive radar imaging.
\newblock In {\em 2007 IEEE Radar Conference}, pages 128--133, April 2007.

\bibitem{zheng2017}
L.~{Zheng} and X.~{Wang}.
\newblock Super-resolution delay-doppler estimation for ofdm passive radar.
\newblock {\em IEEE Transactions on Signal Processing}, 65(9):2197--2210, May
  2017.

\bibitem{YD_zhang2013}
Y.~D. Zhang, M.~G. Amin, and B.~Himed.
\newblock Sparsity-based doa estimatioin using co-primes array.
\newblock {\em IEEE Int. Conf. Acoustics, Speech, and Signal Processing}, 5
  2013.

\bibitem{gribonval2001}
Rémi Gribonval.
\newblock Fast matching pursuit with a multiscale dictionary of gaussian
  chirps.
\newblock {\em Signal Processing, IEEE Transactions on}, 49:994 -- 1001, 06
  2001.

\bibitem{figueras2006}
R.~M. {Figueras i Ventura}, P.~{Vandergheynst}, and P.~{Frossard}.
\newblock Low-rate and flexible image coding with redundant representations.
\newblock {\em IEEE Transactions on Image Processing}, 15(3):726--739, March
  2006.

\bibitem{jacques2008}
Laurent Jacques and Christophe Vleeschouwer.
\newblock A geometrical study of matching pursuit parametrization.
\newblock {\em Signal Processing, IEEE Transactions on}, 56:2835 -- 2848, 08
  2008.

\bibitem{winkler2007}
V.~Winkler.
\newblock Range doppler detection for automotive fmcw radars.
\newblock pages 166--169, 11 2007.

\bibitem{lutz2014}
S.~{Lutz}, D.~{Ellenrieder}, T.~{Walter}, and R.~{Weigel}.
\newblock On fast chirp modulations and compressed sensing for automotive radar
  applications.
\newblock In {\em 2014 15th International Radar Symposium (IRS)}, pages 1--6,
  June 2014.

\bibitem{feuillen2016}
T.~{Feuillen}, A.~{Mallat}, and L.~{Vandendorpe}.
\newblock Stepped frequency radar for automotive application: Range-doppler
  coupling and distortions analysis.
\newblock In {\em MILCOM 2016 - 2016 IEEE Military Communications Conference},
  pages 894--899, Nov 2016.

\bibitem{zubair2013}
S.~Zubair and W.~Wang.
\newblock Tensor dictionary learning with sparse tucker decomposition.
\newblock pages 1--6, 07 2013.

\bibitem{fang2015}
Y.~Fang, B.~Huang, and J.~Wu.
\newblock 2d sparse signal recovery via 2d orthogonal matching pursuit.
\newblock {\em Science China Information Sciences}, 55, 01 2011.

\bibitem{candes2008}
E.~J. {Candes} and M.~B. {Wakin}.
\newblock An introduction to compressive sampling.
\newblock {\em IEEE Signal Processing Magazine}, 25(2):21--30, March 2008.

\bibitem{azodi2016}
H.~Azodi, C.~Koenen, U.~Siart, and T.~Eibert.
\newblock Empirical discretization errors in sparse representations for motion
  state estimation with multi-sensor radar systems.
\newblock pages 1--4, 04 2016.

\bibitem{tang2013}
G.~{Tang}, B.~N. {Bhaskar}, P.~{Shah}, and B.~{Recht}.
\newblock Compressed sensing off the grid.
\newblock {\em IEEE Transactions on Information Theory}, 59(11):7465--7490, Nov
  2013.

\bibitem{mishra2014}
K.~V. {Mishra}, M.~{Cho}, A.~{Kruger}, and W.~{Xu}.
\newblock Super-resolution line spectrum estimation with block priors.
\newblock pages 1211--1215, Nov 2014.

\bibitem{traonmilin2018}
Y.~Traonmilin and J-F. Aujol.
\newblock The basins of attraction of the global minimizers of the non-convex
  sparse spikes estimation problem.
\newblock {\em ArXiv}, abs/1811.12000, 2018.

\bibitem{simoncelli2011}
C.~{Ekanadham}, D.~{Tranchina}, and E.~P. {Simoncelli}.
\newblock Recovery of sparse translation-invariant signals with continuous
  basis pursuit.
\newblock {\em IEEE Transactions on Signal Processing}, 59(10):4735--4744, Oct
  2011.

\bibitem{knudson2014}
K.~Knudson, J.~Yates, A.~Huk, and J.~Pillow.
\newblock Inferring sparse representations of continuous signals with
  continuous orthogonal matching pursuit.
\newblock {\em Advances in neural information processing systems}, 27, 04 2015.

\bibitem{liu2010}
Y.~Liu, H.~Meng, G.~Li, and X.~Wang.
\newblock Velocity estimation and range shift compensation for high range
  resolution profiling in stepped-frequency radar.
\newblock {\em Geoscience and Remote Sensing Letters, IEEE}, 7:791 -- 795, 11
  2010.

\bibitem{bao2014}
H.~Bao.
\newblock The research of velocity compensation method based on range-profile
  function.
\newblock {\em International Journal of Hybrid Information Technology},
  7:49--56, 03 2014.

\end{thebibliography}
\end{small}

\end{document}